\documentclass[a4paper]{jpconf}
\usepackage{graphicx}
\begin{document}
\title{Hydrodynamic models of particle production}

\author{Piotr Bo\.zek$^{1,2}$, Wojciech Broniowski$^{2,3,4}$,
 Giorgio Torrieri$^{5,6}$}
\address{$^1$ AGH University of Science and Technology, Faculty of Physics and Applied Computer Science, al. Mickiewicza 30, 30-059 Krakow, Poland}
\address{$^2$ Institute of Nuclear Physics PAN, PL-31342 Krak\'ow, Poland}
\address{$^3$ Institute of Physics, Jan Kochanowski University, PL-25406~Kielce, Poland}
\address{$^4$ CNRS, URA2306, Institut de Physique Th\'eorique de Saclay, F-91191 Gif-sur-Yvette, France}
\address{$^5$ FIAS, J.W. Goethe Universit\"at, Frankfurt A.M., Germany}
\address{$^6$ Pupin Physics Laboratory, Columbia University, 538 West 120$^{th}$ Street NY 10027, USA}

\begin{abstract}
Viscous hydrodynamics gives a satisfactory description of the transverse 
momentum spectra, of the elliptic and triangular flow, and of the femtoscopic 
correlations for  particles produced in relativistic heavy-ion collisions. 
On general grounds, a similar collective behavior has been predicted
for proton-lead (p-Pb) collisions at the LHC. We present results of the 
hydrodynamics calculation of the elliptic and triangular  flow in p-Pb.
We discuss the mass dependence of flow coefficients and of
 the average transverse momentum for identified particles.
\end{abstract}

\section{Hydrodynamic model}

The observation of the  jet quenching and of the strong azimuthally asymmetric flow shows that 
in a high energy heavy-ion collision at ultrarelativistic energies a dense and hot droplet of matter is formed. 
Event-by-event hydrodynamic calculations  \cite{Schenke:2010rr,Petersen:2010cw,Gardim:2011xv,Bozek:2012fw,Qiu:2011hf,Pang:2012he} 
are used to provide the response
of the collective flow to the deformations of the fluctuating initial state.
The eccentricity and triangularity of
the initial density result in the formation of the elliptic and triangular flow in the spectra 
\cite{Gardim:2011xv,Niemi:2012aj}. Not only the average,  but the whole 
distribution of the harmonic flow coefficients $v_n$ 
\cite{Aad:2013xma}
can be reproduced \cite{Gale:2012rq} for the IP-Glasma initial conditions. The 
 event-by event distribution of the hydrodynamic 
response shows
that  $v_2$ and $v_3$ follow the initial distribution of $\epsilon_2$ and $\epsilon_3$,  with nonlinearities in the response appearing for the 
harmonics $v_4$ and higher. Non-flow correlations between particles with soft momenta can be included
 in the simulation giving correlations between unlike-sign particles mainly at
small relative  angle and relative pseudorapidity  \cite{Bozek:2012en}.
The odd and even components  of the directed flow $v_1$
  as a function of pseudorapidity, appear from the expansion of 
a tilted, fluctuating fireball
\cite{Bozek:2010bi,Csernai:2011gg,Teaney:2010vd,Retinskaya:2012ky}.

The harmonic event-plane for each order can be determined from the initial density
 \cite{Teaney:2012ke,Heinz:2013bua}, while the hydrodynamic evolution that
 involves nonlinear coupling between harmonics
 yields additional correlations between the event planes, in good agreement with experiment
\cite{Heinz:2013bua}. Fluctuations may lead to a decorrelation of the event planes at 
different rapidities
\cite{Bozek:2010vz,Petersen:2011fp,Xiao:2012uw}, a prediction that could be tested experimentally. 
Fluctuations of the volume of the initial fireball 
lead to event-by-event fluctuations of the average  transverses momentum \cite{Broniowski:2009te}, 
in an analogous way as for the response of  the flow harmonics to the fireball anisotropies.
The hydrodynamic model gives predictions for the spectra of identified particles that are in satisfactory
 agreement with experiment at LHC energies
 \cite{Shen:2011eg,Bozek:2012qs,Karpenko:2012yf,Werner:2012xh,Abelev:2013vea}.
Advanced hydrodynamic calculations are used to extract the properties of the quark-gluon plasma
at different temperatures 
\cite{Shen:2011zc,Niemi:2011ix,Bozek:2009dw,Gale:2012rq,Luzum:2012wu} and barionic densities 
\cite{Steinheimer:2012bn,karpenkosqm}. The hard equation of state of the quark-gluon plasma
 leads to a rapid build up of the transverse flow, which implies a strong  dependence of the
 femtoscopic radii on the pion pair momentum \cite{Broniowski:2008vp,Pratt:2008qv}.

The  p-Pb
 collisions  have been proposed as a laboratory to study the initial 
state effects and to obtain reference data for the Pb-Pb collisions  
\cite{Salgado:2011pf}. On the other hand, it has been predicted that
collective expansion of the fireball formed in central p-Pb collisions is 
significant and can lead to observable elliptic and triangular flow
\cite{Bozek:2011if}.
 Experiments with p-Pb 
collisions at the LHC measure two-particle correlation 
function in relative pseudorapidity $\Delta \eta$ and relative azimuthal angle
$\Delta \phi$
\cite{CMS:2012qk,Abelev:2012ola,Aad:2012gla}.  The two-dimensional
 correlation function presents two ridge-like structures, elongated in 
pseudorapidity, with pairs collimated in the same direction (same-side ridge, $\Delta \phi\simeq 0$) and in the opposite direction (away-side ridge, $\Delta \phi \simeq \pi$). The azimuthal structure 
in  the two-dimensional correlation
function can be understood as an effect of the azimuthally
asymmetric  collective flow and of the transverse momentum conservation 
\cite{Bozek:2012gr}, in a similar way as in heavy-ion collisions 
 \cite{Takahashi:2009na,Luzum:2010sp}.
The formation of the 
two ridges is predicted in the color glass condensate approach  as well
\cite{Dusling:2012cg,Dusling:2013oia,Dusling:2012wy}.
In Ref. \cite{Bozek:2011if} it has been proposed to look for collectivity in small systems using 
d-A collisions. In central d-A collisions the density profile in the transverse plane is determined
 by the deuteron wave-function, hence the eccentricity $\epsilon_2$ is large $\simeq 0.5$, which gives  a large
elliptic flow \cite{Bozek:2011if}. The analysis of correlation functions in d-Au collisions at 
RHIC shows a strong $v_2$ component \cite{Adare:2013piz}, but  quantitative conclusions
are difficult due to large non-flow contributions in the low multiplicity environment at the RHIC energies.

\section{Transverse flow in p-Pb interactions}

The multiplicity in central p-Pb collisions is comparable to peripheral Pb-Pb collisions.
The source size can be estimated using the Glauber Monte Carlo model \cite{Bozek:2013uha,Bzdak:2013zma}.
The value of the root mean square radius, 1-2~fm, depends on the details of the energy deposition in 
the model, and grows for more central collisions. 
\begin{figure}[h]
\begin{minipage}{7.cm}
\includegraphics[height=7cm]{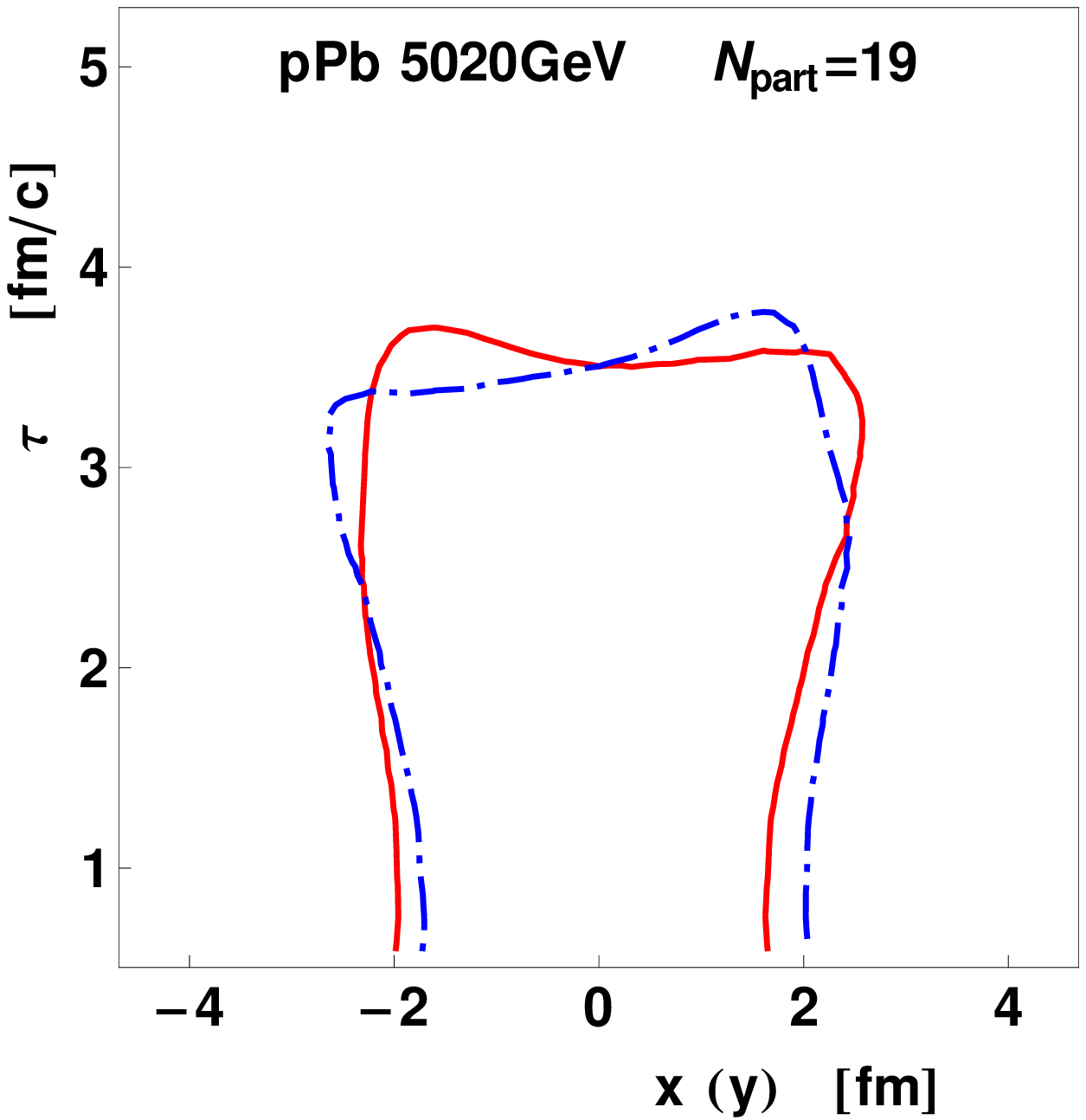}
\end{minipage}\hspace{2pc}%
\begin{minipage}{9.cm}
\includegraphics[height=7cm]{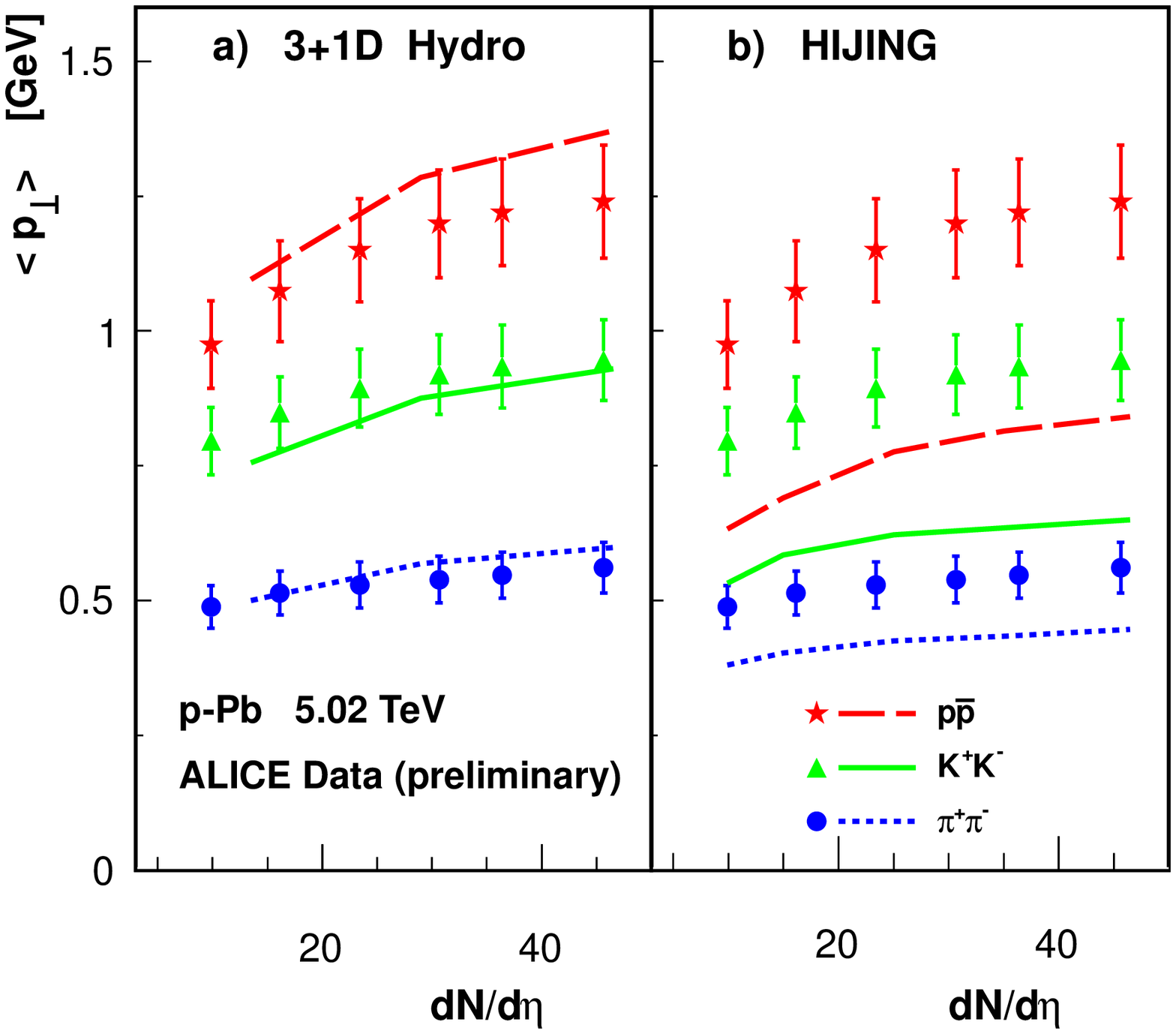}
\end{minipage} 
\caption{\label{fig:pt}  Left: 
The freeze-out 
isotherms in the $x-\tau$ plane (solid) and the $y-\tau$ plane (dashed) in a sample event. Right: 
Mean transverse momentum of identified particles as a function of charged particle density
from hydrodynamics (panel (a)) and HIJING~ (panel (b)), ALICE Collaboration data~\cite{Abelev:2013bla}.}
\end{figure}
The expansion of the large energy density deposited in the small volume generates large
transverse collective flow \cite{Bozek:2013uha,Bzdak:2013zma,Bozek:2013df,Pierog:2013ria}.
In the left panel of Fig. \ref{fig:pt} is shown the freeze-out hypersurface for an event 
with 19 participant nucleons. The transverse size increases during the freeze-out and
 can be measured using interferometry radii. The values of the interferometry radii grow 
as a power of the multiplicity $N^{1/3}$ \cite{Kisiel:2011jg}. The slope of this dependence is 
different for A-A and for p-p collisions. It may suggest that the dynamics of a typical p-p interaction 
is different than in peripheral A-A collisions. The hydrodynamical model predicts for the p-Pb system
  interferometry radii that are close to those corresponding to A-A interactions \cite{Bozek:2013df}.

The average transverse momentum of particles produced in the 
p-p, p-Pb and A-A interactions increases with 
the event multiplicity \cite{Abelev:2013bla,Chatrchyan:2013eya}. For p-p interactions this
increase can be explained as a color reconnection effect \cite{Ortiz:2013yxa}. Modeling the particle
 production in p-Pb and A-A collisions as a superposition of nucleon-nucleon interactions one finds
the average transverse momentum  below the measured value \cite{Bzdak:2013lva,Abelev:2013bla}.
This effect is visible in the right panel of Fig. \ref{fig:pt}, where the HIJING model,
 based on the superposition
of nucleon-nucleon collisions, underpredicts the average transverse momentum.
This difference leaves room for additional collective transverse velocity
  to be generated during the expansion phase in p-Pb collisions. An important characteristic of 
the collective transverse flow is the mass hierarchy. The average transverse momentum from collective flow 
is larger for heavier particles. The data for $\pi$, $K$, and p transverse momenta can
 be reproduced naturally in the hydrodynamic framework \cite{Bozek:2013ska} 
(Fig. \ref{fig:pt} right panel).  The consistency
 of the hydrodynamic calculations with the experimental data validates the collective flow interpretation,
while we note that the mass hierarchy of the average transverse momentum 
can also be understood as coming from geometrical scaling \cite{McLerran:2013oju}.
The transverse momentum spectra in p-Pb collisions contain a soft component coming from the collective 
expansion with statistical emission at freeze-out \cite{Bozek:2011if,Pierog:2013ria,Abelev:2013haa}.
The spectra can be reproduced in the region of intermediate $p_\perp$ in the EPOS LHC model 
\cite{Pierog:2013ria,Abelev:2013haa,Chatrchyan:2013eya}.

\section{Elliptic and triangular flow in p-Pb interactions}

\begin{figure}[h]
\begin{minipage}{7.5cm}
\includegraphics[height=5.4cm]{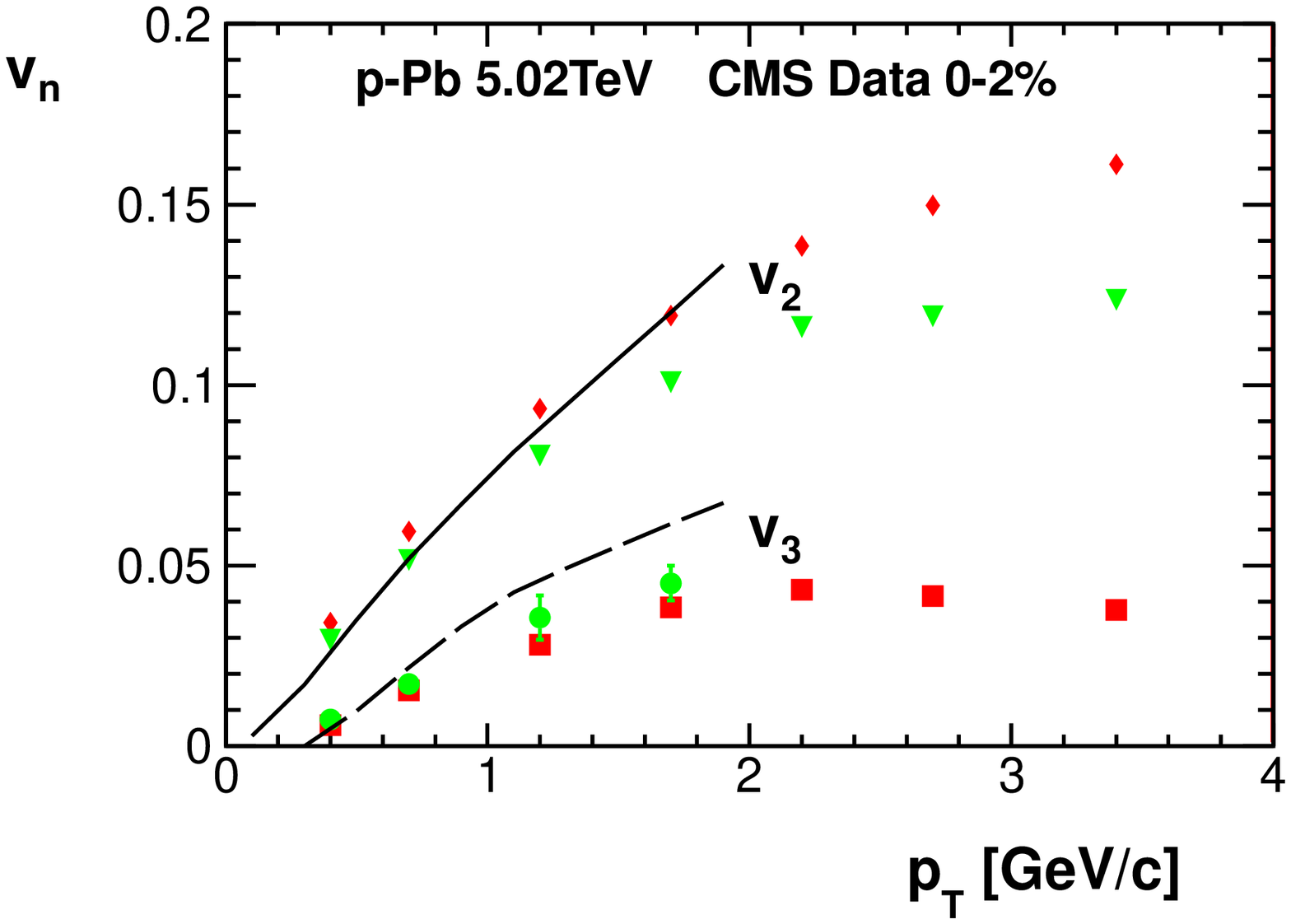}
\end{minipage}\hspace{2pc}%
\begin{minipage}{7.5cm}
\includegraphics[height=5.4cm]{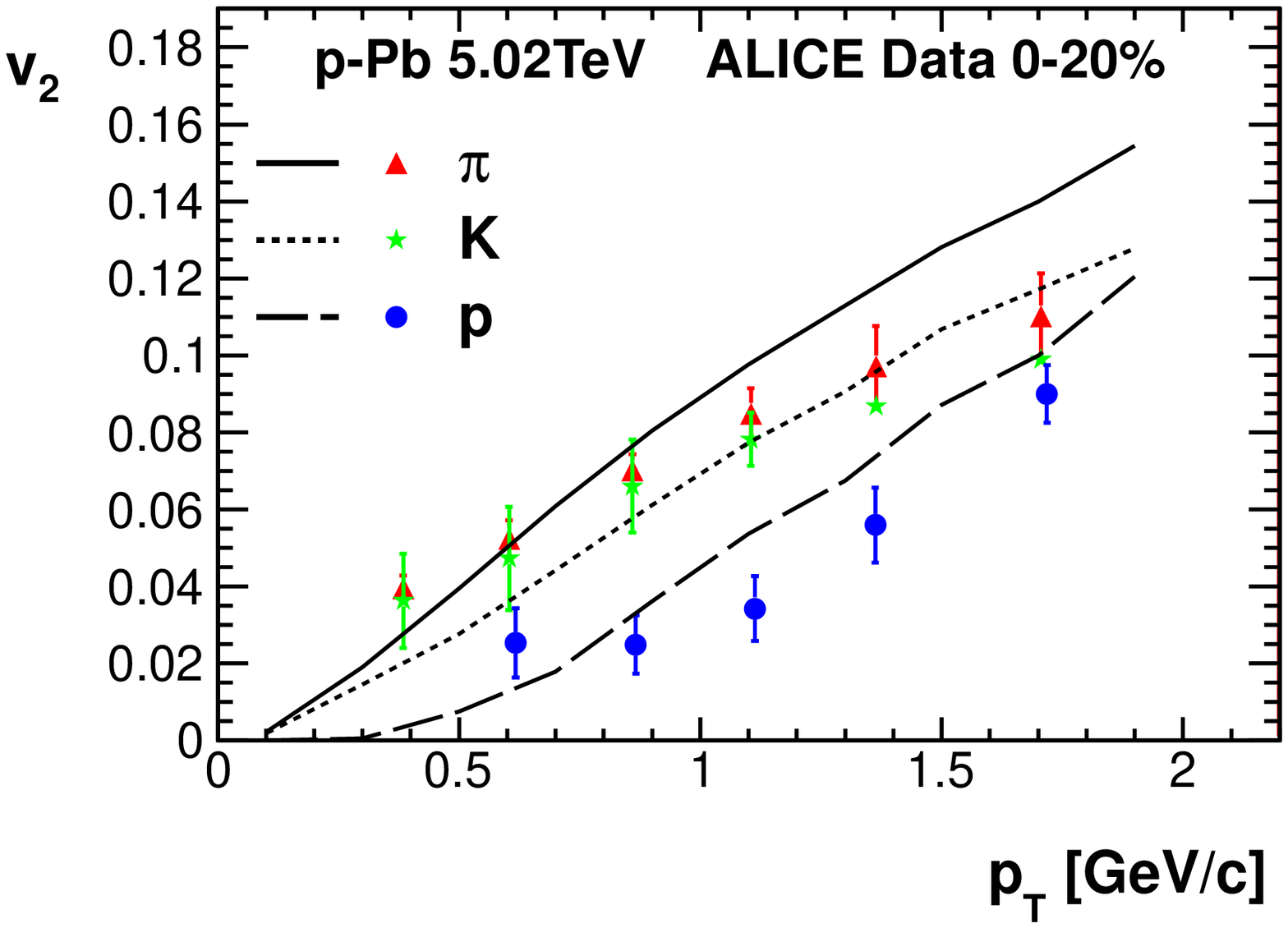}
\end{minipage} 
\caption{\label{fig:v2id} Left:   $v_2$ and $v_3$ for charged 
particles from the hydrodynamic calculation,
 CMS Collaboration data  ~\cite{Chatrchyan:2013nka}. Right:  $v_2(p_\perp)$ for pions, kaons 
and protons from the hydrodynamic model, 
ALICE Collaboration data~\cite{ABELEV:2013wsa}).}
\end{figure}

The most important evidence for the collective expansion in the A-A collisions is the
observation of elliptic and triangular flow. The elliptic flow coefficient $v_2$ has been measured using
2- and 4-particle cumulants in p-Pb collisions
 \cite{Aad:2013fja,Chatrchyan:2013nka}. The data is consistent 
with the predictions of
  the hydrodynamic model \cite{Bozek:2011if,Bozek:2013uha,Qin:2013bha,Werner:2013ipa} 
(Fig. \ref{fig:v2id} right panel). 
In p-Pb interaction the model 
uncertainty on the  value of the initial eccentricity is a more severe problem than for the A-A system
\cite{Bozek:2013uha,Bzdak:2013lva}. The small size and short lifetime of the system 
 make the final results depend significantly on other parameters of the model as well: the 
initial time of the expansion, the freeze-out temperature, the shear viscosity. An additional 
difficulty in comparing to the experimental data is related to the large contribution of non-flow 
correlations that are not implemented in the models. The relativistic viscous
 hydrodynamics can be reliably applied if the velocity gradients in the system are moderate and if the 
mean free path is smaller than the size of the system. In small systems the contribution of 
non-thermal corona is relatively more important, which can reduce the value of the elliptic flow.
These assumptions become more and more 
questionable when applying the model to more peripheral p-Pb collisions. All of the above mentioned 
factors mean that the agreement of the calculation with the data can be at best semi-quantitative.

An important experimental result is the observation of mass splitting of the elliptic flow in
 p-Pb interactions \cite{ABELEV:2013wsa}. 
The smaller value of the elliptic flow coefficient $v_2(p_\perp)$
for heavy particles is a characteristic of the collective elliptic flow. This feature is reproduced
by the calculation \cite{Bozek:2013ska,Werner:2013ipa} (Fig. \ref{fig:v2id} right panel).
The observation of triangular flow in p-Pb interactions is a strong argument for the
 collective expansion scenario. The predictions of the hydrodynamic model, with fluctuating initial 
conditions are consistent with measurements  \cite{Bozek:2011if,Bozek:2013uha,Qin:2013bha}
 (Fig. \ref{fig:v2id} right panel). This means that the model captures correctly the fluctuations of the
initial state and  describes realistically 
 its collective  expansion. The mass splitting for the triangular
 flow $v_3$ in p-Pb collisions
is smaller than for $v_2$, moreover, it is distorted by resonance decays.

\section{Collectivity in p-Pb}

The formation of a fireball of strongly interacting fluid in ultrarelativistic heavy-ion collisions
is well established. The observation of elliptic and triangular flow  in agreement with predictions of
nearly prefect fluid hydrodynamics, the quantitative description of transverse momentum spectra, 
and jet quenching are  strong evidence of  the creation of the 
quark-gluon plasma. 
Quantitatively the hydrodynamic model can be reliably applied in central A-A collisions. Many
 experimental efforts have been devoted to the study the properties of the matter in the fireball
 when changing the system
size and the energy. The data indicate that a collectively expanding system is formed in 
peripheral A-A collisions and at different energies of the RHIC beam energy scan. 

It has been generally expected that in  p-Pb collisions at the LHC, the 
final state 
interactions  would be negligible \cite{Salgado:2011pf}, but the hydrodynamic 
model applied to initial conditions extrapolated to the p-Pb system
 at TeV energies predicted
 a significant collective expansion \cite{Bozek:2011if} that could be observed 
as the elliptic and triangular flow.  Many experimental data from p-Pb interactions are 
consistent with the collective expansion scenario, the double-ridge \cite{Bozek:2012gr}, the
elliptic and triangular flow \cite{Bozek:2011if,Bozek:2013uha,Qin:2013bha,Werner:2013ipa},
 the mass hierarchy of  average transverse momentum \cite{Bozek:2013ska,Pierog:2013ria}, and of the
 elliptic flow \cite{Bozek:2013ska,Werner:2013ipa}. The observed appearance of the flow in
 low multiplicity p-Pb events \cite{Chatrchyan:2013nka} and general arguments suggest that
the p-Pb system can be used as a testing ground for the onset of collectivity in small systems.

\section*{Acknowledgments}
 PB and WB acknowledge the support of the Polish
National Science Centre, grant DEC-2012/06/A/ST2/00390 and PL-Grid infrastructure. GT acknowledges the financial support received from the Helmholtz International
Centre for FAIR within the framework of the LOEWE program
(Landesoffensive zur Entwicklung Wissenschaftlich-\"Okonomischer
Exzellenz) launched by the State of Hesse, and from DOE under Grant No. DE-FG02-93ER40764.

\section*{References}
\providecommand{\newblock}{}

\end{document}